\documentclass[aps,prl,twocolumn,showpacs,superscriptaddress,groupedaddress]{revtex4}  
\usepackage{graphicx}  
\usepackage{dcolumn}   
\usepackage{bm}        
\usepackage{amssymb}   
\usepackage{amsmath}
\usepackage{lipsum} 
\usepackage{braket}
\usepackage{natbib,hyperref}
\usepackage{siunitx}
\usepackage{xcolor}

\newcommand{\Nat}{N_{\text{SPH}}}

\newcommand{\gammaH}{\gamma_{\text{H}}}

\newcommand{\mum}{\text{ }\mu_{\text{m}}}
\newcommand{\mm}{\text{ mm}}
\newcommand{\cm}{\text{ cm}}

\renewcommand{\url}[1]{\href{#1}{Link}}

\begin{document}

\title{A Nanosecond-Resolved Atomic hydrogen Magnetometer}
\date{\today}

\author{Alexandros K. Spiliotis}
\author{Michalis Xygkis}
\author{Konstantinos Tazes}
\author{George E. Katsoprinakis}
\affiliation{Foundation for Research and Technology Hellas, Institute of Electronic Structure and Laser, N. Plastira 100, Heraklion, Crete, Greece, GR-71110}
\affiliation{University of Crete, Department of Physics, Heraklion, Greece}
\author{Dimitrios Sofikitis}
\altaffiliation[Permanent address: ]{University of Ioannina, Department of Physics, Ioannina, Greece}
\affiliation{Foundation for Research and Technology Hellas, Institute of Electronic Structure and Laser, N. Plastira 100, Heraklion, Crete, Greece, GR-71110}
\affiliation{University of Crete, Department of Physics, Heraklion, Greece}
\author{Georgios Vasilakis}
\affiliation{Foundation for Research and Technology Hellas, Institute of Electronic Structure and Laser, N. Plastira 100, Heraklion, Crete, Greece, GR-71110}
\author{T. Peter Rakitzis}
\affiliation{Foundation for Research and Technology Hellas, Institute of Electronic Structure and Laser, N. Plastira 100, Heraklion, Crete, Greece, GR-71110}
\affiliation{University of Crete, Department of Physics, Heraklion, Greece}

\begin{abstract}

We introduce a novel and sensitive ns-resolved atomic magnetometer, which is at least three orders of magnitude faster than conventional magnetometers. The magnetic field dependence of hyperfine beating of high-density spin-polarized H atoms, produced from the rapid photodissociation of HCl gas with sub-ns laser pulses, results in a few nT sensitivity for a spin-projection limited sensor with 10 nl measurement volume after 1 ns measurement time. The magnetometer will allow ultrafast continuous B-field measurements in many fields, including spin chemistry, spin physics, and plasma physics.
\end{abstract}

\maketitle

Sensitive detection of magnetic fields plays an important role in several fields, including biomedicine, materials science, security screening, geophysical and space surveys \cite{book:MagneticSensors}, as well as fundamental physics measurements, such as axion searches \cite{AxionSearches}.
In addition to the requirement for high sensitivity, numerous applications demand magnetic field detection with high temporal resolution. For instance, ultrafast magnetometry can be used as a sensitive probe of the pathways and kinetics of chemical reactions involving the production of electron or nuclear spin polarization, such as in photo-chemically induced dynamic nuclear polarization (photo-CIDNP) \cite{Gnezdilov2003, Mompean2018, Morozova, KleinSeetharaman}.
Similarly, studies of magnetodynamics can shed light on dynamical processes in magnetic materials \cite{Kerr_Magnetometry_Science, Kerr_Magnetometry_studies} and plasmas \cite{MagneticReconnection_RMP}, enabling the development of new technologies.
For imaging purposes or for mapping the magnetic field gradients, an important quality is the length scale of the magnetometer, as,  in general, small-sized sensors allow for high spatial resolution.

A plethora of magnetometers has been developed to address the challenges in magnetic field sensing (see for example \cite{book:HighSensitivityMagnetometers} and references therein).
Atomic magnetometers based on optically polarized alkali-metal atoms have realized the highest sensitivities \cite{OpticalMagnetometry}, albeit at low bandwidths, typically up to a few tenths of kHz \cite{JimenezHighBandwidthMagnetometer}. Superconducting quantum interference devices (SQUIDs) have demonstrated similar sensitivities. Although in principle their frequency response can extend up to a few GHz, practical considerations related to the electronic circuitry for the signal readout reduce their bandwidth to the MHz range \cite{Fagaly_SQUIDS_Review}. Inductive-coil sensors feature high detection-bandwidth and broad applicability to different spin species and environments, but they present relatively low sensitivity compared with other state of the art magnetometers (see however \cite{InductionCoilsBecomeQuantunm} for an ultra-sensitive, quantum limited inductive sensor). Optical approaches based on the Kerr effect have realized magnetometry in the picosecond temporal regime; nevertheless, their inherent sensitivity is limited and they are mainly used to probe rather large fields from magnetic materials \cite{Kerr_Magnetometry_Science, McCord_2015}. Nitrogen-vacancy defects in diamond offer an attractive platform for high spatial resolution\cite{Budker_NVD}. Yet, the demonstrated sensitivity does not rival alkali-atomic magnetometers or SQUIDs and elaborate dynamic decoupling techniques are required to realize broadband magnetometry.

Here, we present a novel atomic hydrogen magnetometer that can measure magnetic fields with high sensitivity and nanosecond time resolution. The sensitive detection results from realizing large spin densities, many orders of magnitude higher than conventional alkali-metal atomic sensors, while the temporal resolution derives from performing magnetometry at the hyperfine coherences of hydrogen(hyperfine frequency of 1.42~GHz and 0.327~GHz in H and D, respectively). The proposed magnetometer operates at room temperature, is technically simple to use and can be adjusted to sense sub-micrometer length-scales.

High density spin-polarized hydrogen (SPH) is produced by photodissociating hydrohalide gas with a circularly-polarized UV laser pulse, by exciting a well-defined dissociative electronic state of the molecule (see \cite{Rakitzis_Science, Sofikitis_HClHBr_SPH, Rakitzis_ChemPhysChem} for details).
In hydrohalide photolysis, the fragments acquire a spin-orientation which is correlated to the spin of the dissociating photon, and their final polarization is limited by the dissociation path and by the random direction of the hydrohalide molecular bond with respect to the laser propagation direction. Under readily achieved experimental conditions concerning the laser wavelength and intensity, the photodissociation probability can be near unity, thereby the density of the atomic fragments can be made similar to the initial density of the parent molecule.
With this process SPH densities of $10^{19} \textrm{ cm}^{-3}$ and up to 40\% polarization have been demonstrated for isotropic molecular bonds \cite{Sofikitis_HClHBr_SPH,Sofikitis_HighDensity,Spiliotis_LSA}, whereas up to 100\% is possible if the bonds are aligned along the photodissociation-laser polarization axis \cite{Sofikitis_DI}.

Crucially for the operation of the magnetometer, photolysis is predominantly an electronic process occurring within sub-picosecond time-scales, during which the nuclear spin degrees of freedom are effectively frozen \cite{Sofikitis_HighDensity}. For hydrogen halides, photodissociation with a sufficiently fast laser pulse typically generates fragments with electronically polarized spins, while the nuclear spins remain in their thermal unpolarized state.

We consider SPH produced from HCl photolysis with a 150 ps laser pulse at 213~nm, which has favorable dissociating channels for producing large spin orientation \cite{Spiliotis_LSA}.
Immediately after the photodissociation (and assuming that the HCl bonds have been aligned parallel to the photodissociation polarization direction), half of the generated H atoms are in the state $\left | \psi_0 \right \rangle= \left |m_s =  \sigma/2, m_I = \sigma/2 \right \rangle_{\text{u}}= \left| F=1, m_F = \sigma \right \rangle_{\text{c}}$ and half of the polarized atoms are in the $\left| \psi_1 \right \rangle= \left |m_s = \sigma /2 , m_I = -\sigma/2 \right \rangle_{\text{u}} = \frac{1}{\sqrt{2}} \left( |F=1,m_F=0\rangle_{\text{c}}-\sigma|F=0,m_F=0\rangle_{\text{c}}\right)$. Here, $F$ is the total spin (sum electronic and nuclear spin) quantum number, $m_s$, $m_I$ and $m_F$ are respectively the electronic, nuclear and total spin projection along the quantization axis, the subscripts u and c refer to the uncoupled and coupled angular momentum basis respectively, $\sigma$ is the helicity of the photons (+1 or -1 for $\sigma$+ and $\sigma$- respectively), and the quantization axis is taken to be along the laser propagation direction.
Unlike $\left| \psi_0 \right \rangle$, the state $\left | \psi_1 \right \rangle$ is not an eigenstate of the hyperfine interaction. Atoms prepared initially in $\left| \psi_1 \right \rangle $ experience hyperfine quantum beats, causing the electron and nucleus to exchange spin at the hyperfine frequency. In the absence of a magnetic field this frequency corresponds to the 21-cm hydrogen line at $f_0 \approx 1.42$~GHz.

A magnetic field parallel to the dissociation axis does not affect to first order the hyperfine coherences resulting from the photodissociation $\left|0,0 \right>_{\text{c}} \leftrightarrow \left|1,0 \right>_{\text{c}}$.
However, the presence of a magnetic field perpendicular to the pumping direction modifies the hyperfine oscillation and induces a net precession of the spins. Taking the direction of magnetic field as the quantization axis, the evolution of $\left| \psi_1 \right \rangle$ can be written in the form (ignoring a global phase factor and small amplitude terms of order $\gammaH B/\omega_0$):
\begin{equation}
\left| \psi(t)_1 \right \rangle=  \frac{1}{2} \left [ e^{-\imath \omega_{+} t} \left|1, 1 \right \rangle_{\text{c}}' +  e^{-\imath \omega_{-} t} \left| 1, -1 \right \rangle_{\text{c}}'-\imath \sigma \sqrt{2} \left|0, 0 \right \rangle_{\text{c}}' \right], \label{eq:StateEvolution}
\end{equation}
where $t$ is the time after the photodissociating pulse, the apostrophe distinguishes the quantization axis and the angular frequencies are given by the Breit-Rabi formula:
\begin{equation}
\omega_{\pm} \approx \omega_0 \pm \gammaH B + \left(\frac{\gammaH B}{\omega_0}\right)^2, \label{eq:HyperfineFrequencies}
\end{equation}
In the above equation, $\omega_0 = 2 \pi f_0$, $\gammaH \approx 2 \pi \times 1.4$~MHz/G is the gyromagnetic ratio of total angular momentum for atomic H, $B$ is the magnetic field, and we have ignored the small nuclear gyromagnetic Land\'{e} factor.
Information about the magnetic field can be deduced by monitoring the hyperfine dynamical evolution of the electron spin. In the case of H this is most easily performed with an inductive pickup coil, which measures the time-dependent ensemble magnetization, determined to a good approximation by the hydrogen electron spin.

Spin decoherence in SPH limits the performance of the magnetometer. The spin decay mechanism originates from depolarizing collisions of H with the particles in the ensemble, which include the photofragments and their secondary side-products from chemical reactions, as well as the parent molecules in the case of partial photodissociation (see \cite{Spiliotis_LSA} and references therein). The impact of collisions to spin-relaxation depends on the density, the conditions of photolysis, and the halide species  and a corresponding optimization should be performed according to the requirements for magnetometry. This optimization may involve compromising the SPH density in order to achieve larger coherence times as discussed in \cite{Spiliotis_LSA}.

A note about the polarization of the halide fragment is due here.
Photodissociation also polarizes the electron spin in the halide atoms causing analogous to the hydrogen hyperfine and magnetization oscillations. However, these coherences are a few orders of magnitude shorter-lived, have not yet been observed\cite{Sofikitis_HighDensity}, and do not contribute to the magnetometer signal after a small transient time on the sub-ns timescale.

\begin{figure*}[h]
\includegraphics[width=\textwidth]{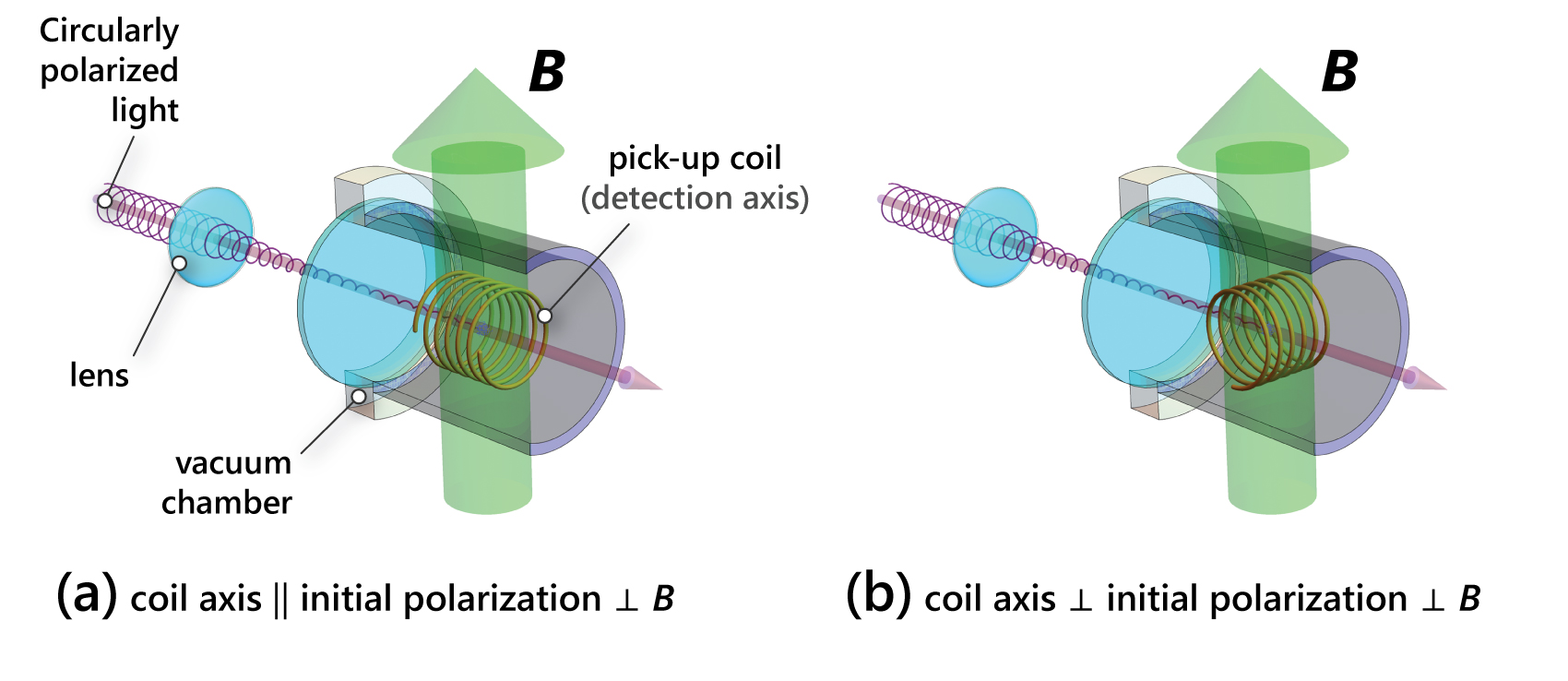}
\caption{(a) Experimental setup schematic for the proof of principle demonstration of the magnetometer. A 150~ps, circularly-polarized laser pulse (red arrow), is focused inside a high density HCl gas cell. A pickup inductive-coil, with its axis parallel to the pumping direction, monitors the ensemble magnetization, while a magnetic field (green arrow) perpendicular to the coil and pumping light is applied; (b) Proposed configuration for sensitive detection of time varying magnetic fields. Compared to (a) the detection axis is perpendicular to both the magnetic field and the light pulse.} \label{fig:ExperimentalSetup}
\end{figure*}

We performed an experiment with a DC magnetic field to demonstrate the magnetometer operation. The experimental setup is based on the one described in \cite{Sofikitis_HighDensity}.
Briefly, a circularly polarized 150~ps laser pulse at 213~nm was focused inside a cell filled with 2~bar of high purity (99.9$\%$) HCl, as shown in Fig.~\ref{fig:ExperimentalSetup}a.
A 4~mm long coil, 4.5 turns, radius of 1~mm and oriented with its axis along the laser pulse propagation, was employed to measure the time-dependent electron magnetization of the SPH atoms. The pickup coil signal was amplified by an RF amplifier and was recorded by a fast-sampling oscilloscope. The time-constant associated with the induction detector was much smaller than the sampling time, so that the recorded signal was directly proportional to the time derivative of the magnetic flux.

The pumping laser beam was weakly focused to a spot of approximately 100~$\mu$m in diameter, located at the center of the coil. For the energy of the laser pulse (3~mJ), the realized intensities were sufficient to dissociate only a small fraction (about 2\%) of the HCl molecules inside the beam volume per pulse. Taking into account the HCl absorption cross section and the photodissociation efficiency we estimate an average SPH density of $10^{18} \text{ cm}^{-3}$ over the coil volume \cite{SpiliotisThesis}. Even though this is about 2\% of the parent molecule density, it is more than three orders of magnitude higher than the densities in alkali-metal atomic magnetometers \cite{JimenezHighBandwidthMagnetometer}. For this proof of principle experiment, we traded off SPH density for increased coherence times, as the spin-relaxation cross section for H-H and H-Cl collisions is much larger than that for H-HCl collisions.

We operated the magnetometer in an unshielded environment. This made the pick-up coil and the cables susceptible to electromagnetic interference, originating mainly from discharges in the laser unit. We filtered this noise by subtracting two consecutive measurements acquired with opposite spin orientations in the SPH ensemble. For this, the helicity of the dissociating laser pulse was reversed every laser shot, using a photoelastic modulator (PEM) and an appropriate electronic circuit to phase lock the low frequency laser emission rate with the higher PEM modulation frequency \cite{SpiliotisThesis}.

A static transverse magnetic field was applied using two permanent round magnets (Neodymium grade N42, 10~mm diameter and 2~mm thickness) placed astride the measurement region with their magnetization vectors aligned. The magnets were attached on translation stages, which allowed us to control their separation. This way, a homogeneous, precisely tunable magnetic field, ranging from 10~G to 150~G was applied. A Hall probe was used to calibrate the magnetic field versus magnet separation at low fields, while at higher fields the field was estimated from extrapolation.

\begin{figure*}[h]
\includegraphics[width=\textwidth]{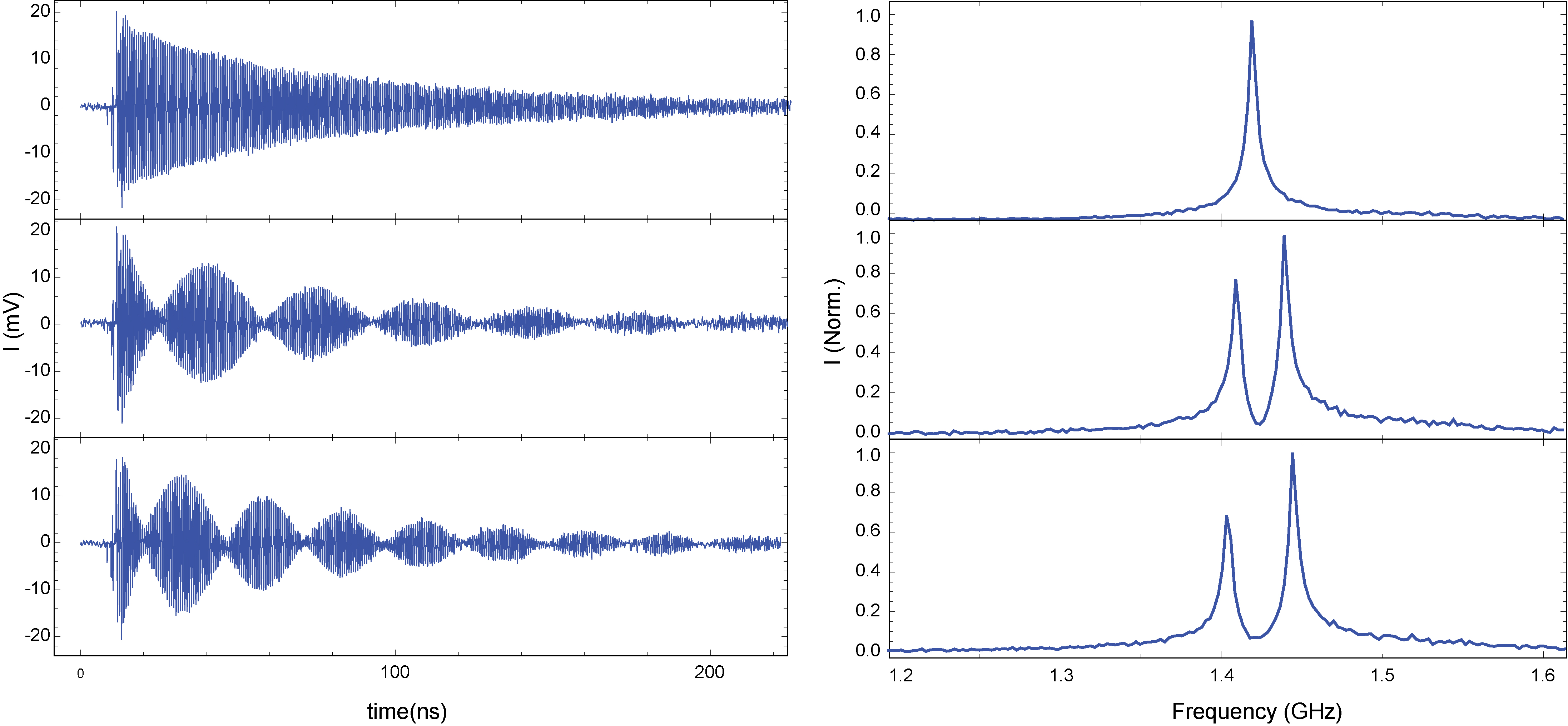}
\caption{Measured signal in the time domain (left) and the corresponding amplitude of Fourier transform (right) for different values of magnetic fields.} \label{fig:data}
\end{figure*}

The measured signals, averaged over 50 repetitions, and the corresponding Fourier transforms for various magnetic fields are shown in Fig.~\ref{fig:data}. The photodissociation pulse occurs at time $t=0$. After a fast transient, which decays at sub-ns timescales, the oscillating signal reflects the H hyperfine coherences damped by collisions. For zero magnetic field ($B=0$) a single frequency at $f_0$ appears in the Fourier spectrum.
A non-zero magnetic field, transverse to the pumping-axis, modifies the frequency of hyperfine coherence and two Fourier peaks appear in the signal, which can be found  from Eq.~\ref{eq:HyperfineFrequencies}.

\begin{figure*}[h]
\includegraphics[width=\textwidth]{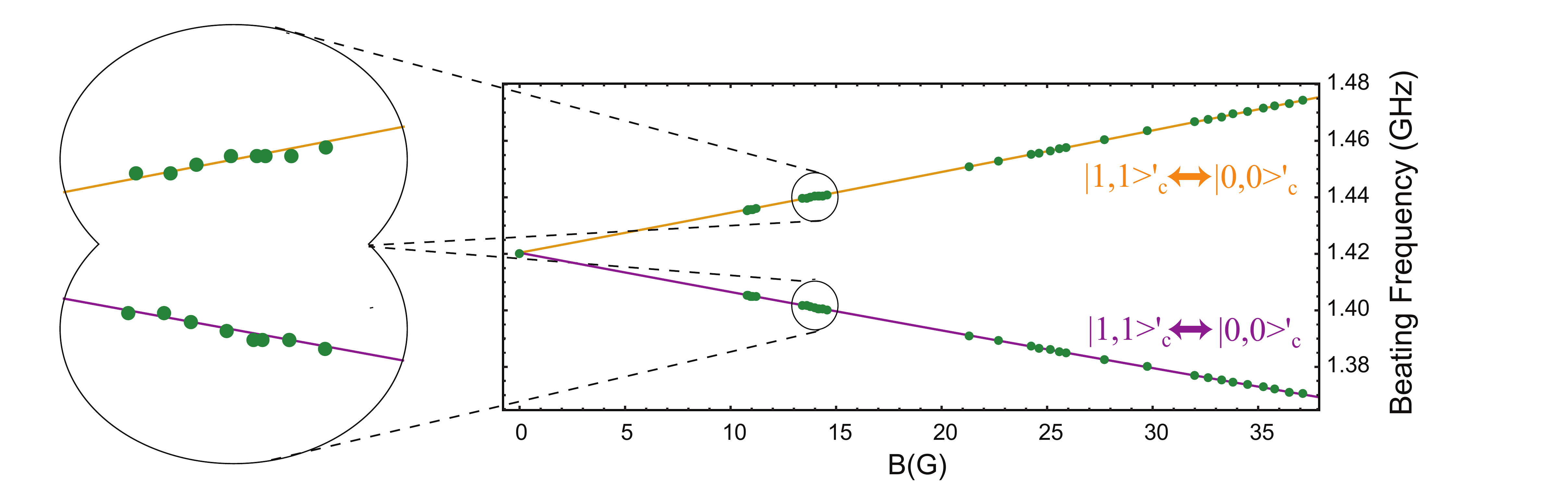}
\caption{Measured hyperfine frequencies (green dots) versus applied magnetic field. The solid lines are the predictions of Eq.~\ref{eq:HyperfineFrequencies}. At every magnetic field two frequencies appear, corresponding to diffe	rent hyperfine coherences as explained in the text.
A zoomed-in plot of the data in the region around 2~G is shown on the left.
The frequency uncertainty (1 sigma) is approximately 40~kHz, smaller than the size of the points.} \label{fig:FreqVsB}
\end{figure*}

In Fig.~\ref{fig:FreqVsB} the measured frequencies are plotted versus the applied magnetic field as quantified from an independent calibration with the Hall probe. The peak frequencies were found using an algorithm which fits a Lorentzian curve to the three consecutive points with the highest amplitudes in the discrete Fourier transform \cite{SpiliotisThesis}.
The agreement between the frequency peaks and the prediction of Eq.~\ref{eq:HyperfineFrequencies} is excellent within the measurement resolution, demonstrating that the magnetic field can be estimated from the Fourier peaks. By repeating the measurement multiple times we estimate that the frequency statistical uncertainty (standard deviation) was around 40~kHz corresponding to approximately 30~mG sensitivity after 50 iterations. In this case, the sensitivity was limited by the poor resolution in Fourier transform, the electronic noise from an unoptimized electric circuitry and from discharge noise originating from pumping the photodissociating laser.

The magnetometer that relies on the frequency content of Fourier power spectrum is not appropriate for characterizing a time-varying field. For such an estimation, the phase evolution of the signal should also be taken into account. In addition, high resolution detection requires first-order sensitivity to the magnetic field.
The configuration that satisfies these conditions is shown in Fig.~\ref{fig:ExperimentalSetup}b: the dissociation/pumping axis, the magnetic field direction and the detection axis are allperpendicular to each other.
With this scheme, the signal $\mathcal{E}$ (electromotive force in the detection coil) at time $t$ after the optical pumping can be approximated to be (see \cite{tazes2021magnetometry} for details):
\begin{equation}
\mathcal{E}(t) =   \mathcal{G} \omega_0 \mathcal{N} \mu_{\text{B}} \Nat  e^{-t/T_2} \sin \left[ \gamma_H \int_0^{t} B(t') dt'  \right] \sin \left( \omega_0 t \right), \label{eq:FastSensitiveMagnetometer}
\end{equation}
where $\mathcal{G}$ is a geometrical factor that relates the magnetic field from the spins to the flux in the region of the coil,  $\mathcal{N}$ is the number of turns per unit length of the coil, $\mu_{\text{B}}$ is the Bohr magneton,  $\Nat$ the number of SPH atoms in the detection region, and $T_2$ is the hyperfine coherence time.
The above equation is an approximation in the limit where the magnetic field evolves much slower compared to the hyperfine frequency and when $\omega_0 \gg \left( 1/T_2, \gammaH B \right)$. It also assumes an inductor detection time constant much shorter than the hyperfine period and neglects dipolar-interactions between the hydrogen spins.
It can be seen that the information about the magnetic field appears as amplitude modulation of the carrier wave at the hyperfine frequency $\omega_0$. As a result, the evolution of magnetic fields at timescales longer than $1/\omega_0$ should be faithfully captured with the proposed magnetometer as long as it is completed before significant spin-depolarization occurs.

\begin{figure*}[hb]
\includegraphics[width=\textwidth]{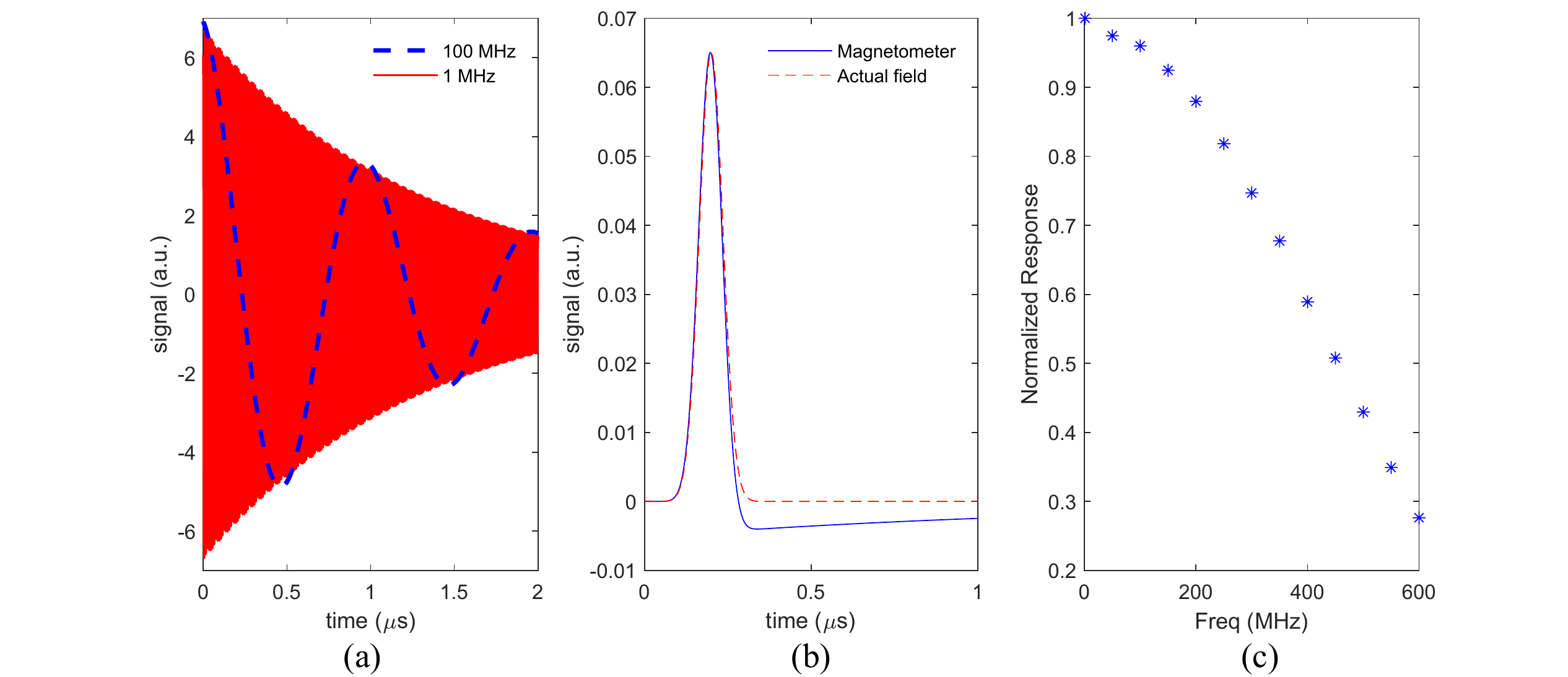}
\caption{Numerical simulation assuming atomic spin decay time 0.75~$\mu$s. The optical pumping axis, magnetic field direction and detection line are all perpendicular to each other as depicted in Fig.~\ref{fig:ExperimentalSetup}b. Plots (a) and (b) were produced in the same (arbitrary) vertical units. (a) $B(t) = B_{0} cos(2\pi f
t)$, where $ B_{0} = 100 \text{ } \mu\text{G}$ and $f=1$~MHz (blue dashed line) or $f=100$~MHz (red solid line). (b) $B(t) = B_{0} e^{-{\frac{(t-t0)^{2}}{\Delta t^{2}}}}$ ,where $\Delta t = 50 \text{ ns}$, $t_{0} = 200 \text{ ns}$ and $B_{0} = \text{ } 100 \mu\text{G} $. (c) Relative magnetometer response to a cosinusoidal magnetic field as a function of the oscillation frequency. The response, characterized by the maximum oscillation amplitude in the readout signal, is normalized to the DC response.} \label{fig:AC_simulation}
\end{figure*}

In Fig.~\ref{fig:AC_simulation} we show the results of simulations for time-varying magnetic waveforms:
(a) is for cosine waves at 1~MHz and 100~MHz, while (b) is for a Gaussian magnetic pulse with duration (Gaussian RMS width) $50$~ns, centered at $200$~ns after the pumping pulse. The same magnetic field amplitude (maximum value) was considered for all the cases.
In all the simulations the depolarization time was taken to be $0.75$~$\mu$s. The magnetic field was estimated by first demodulating the signal at frequency $\omega_0$ and then taking the time derivative of the demodulation output. For the demodulation we implemented a simple low pass filter, where the signal was multiplied by $\cos \left( \omega_0 t\right)$ and the resultant was integrated over a single period $2 \pi/f_0$. Subfigures (a) and (b) are plotted in the same (arbitrary) units for comparison. The response in the two cosine waves in (a) is similar despite a two orders of magnitude difference in frequencies. In (b) the retrieved Gaussian pulse (solid blue line) approximates well the actual Gaussian waveform (dashed red line), except at the larger time-scales, where the spin-depolarization distorts the recovered waveform. In sub-figure (c), the response to a cosinusoidal magnetic field is plotted as a function of frequency. Here, the response is taken to be the maximum oscillation amplitude in the readout signal; in the examined frequency range, this maximum occurs at time scales much shorter than the decay time and is therefore not affected by the decay. The magnetometer bandwidth, defined as the frequency where the response has decreased by a factor of two, is approximately 450~MHz.

The detection of magnetic fields is fundamentally limited by the spin-projection noise \cite{OpticalMagnetometry}, originating from the Heisenberg uncertainty for spins. Considering the experimental setup shown in Fig.~\ref{fig:ExperimentalSetup}b, the effect of spin noise on the magnetometer depends on the magnetic waveform. We consider a magnetic field of the form: $B(t) = B_0 \mathcal{K}(t)$, where $B_0$ signifies the amplitude of magnetic field to be estimated from the magnetometer signal and $\mathcal{K}$ is an arbitrary function of time $t$ assumed to be known. By adopting a $\chi^2$ minimization, it can be shown \cite{tazes2021magnetometry} that the spin-projection noise limit for a single measurement run is given by:
\begin{widetext}
\begin{equation} \label{sensitivity}
	\delta B_0^2= \frac{2}{\Nat} \frac{\int_0^{T_m} \int_0^{T_m} dt dt'   e^{-  \left( t +t' \right)/T_2 } e^{- |t-t'|/T_2}\int_0^{t} \mathcal{K}(x) dx \int_0^{t'} \mathcal{K} (x') dx' }{ \left[ \gammaH \int_0^{T_m} dt  e^{-2 t/T_2} \left( \int_0^{t} \mathcal{K} (x) dx\right)^2 \right]^2},
\end{equation}
\end{widetext}
where $T_m$ is the measurement time after the photodissociation.
For concreteness we consider the case of a cosinusoidal signal ($\mathcal{K} = \cos \omega t$, where $\omega$ is the angular frequency) and the conditions described above: SPH density $10^{18} \cm^{-3}$, measurement volume $(100 \mum)^2 \times 1 \mm = 10 \text{ nl}$ and $T_2 \approx 100$~ns. Such a projection noise limited magnetometer presents sensitivity on the order of a few nanoTesla after measurement time of 1~ns. For $T_m \sim T_2$ the sensitivity is better than 200~pT/pulse; in this case,  with a pulse repetition rate of 1 MHz, the projected sensitivity becomes better than 200~fT/$\sqrt{\text{Hz}}$.

The magnetometer is also inflicted from thermal and quantum fluctuations in the detection coil and associated electric circuit. The impact of this noise source depends on the geometric arrangement of the spin-ensemble and the pick-up inductor.
Furthermore, for non-spherical measurement volumes the dipolar interactions between the spins impair the magnetometer operation.
We defer the consideration of these issues to a later study.

The presented magnetometer is amenable for miniaturization. The measurement lengthscale, determined by the size of the photodissociation beam, can be readily adapted to the sub-micrometer region. Due to the high densities that can be achieved, the number of polarized atoms remains sufficient for high sensitivities even with such small measurement regions. 
Employing large spin-polarized density also relaxes the requirements on the induction coil. Unless high sensitivities are pursued or the measurement is performed from a very small region, moderate coil quality factor and room-temperature, standard electric circuitry should be adequate to detect the response of the polarized spins to the magnetic field.
Another attractive feature is the large dynamic range and the capacity to operate in relatively high magnetic field backgrounds. The upper detection limit $B_{\text{l}}$ is set by the condition $\gammaH B_\text{l} \ll \omega_0$, which for SPH corresponds to $B_\text{l}$ in the hundreds of Gauss range.

Contrary to the conventional alkali-metal atomic magnetometers, the proposed scheme does not rely on a magnetic resonance condition. Therefore the detection bandwidth is not determined by the decoherence rate, but only from the hyperfine frequency.
In the new scheme the spin-state preparation results from a single-step mechanism (photodissociation of the parent molecule) and does not involve several excitation cycles or the application of magnetic pulses to create coherences.
Unlike magnetometry with nitrogen-vacancy centers in diamond, the dense spin ensemble does not suffer from inhomogeneous broadening and high sensitivities can be achieved without the use of dynamic decoupling techniques.

In conclusion, we have described a novel type of atomic magnetometer which combines sensitive detection with high temporal, nanosecond resolution.
The magnetometer exploits molecular photodissociation in dense hydrohalide gases to produce magnetically sensitive hyperfine coherences in the atomic H fragment.
This approach can be applied to different alkali-metal halides with potential advantage in the time resolution. For instance, operation with Cs instead of H will extend the range of application to the sub-ns regime.
This magnetometer holds great promise for characterizing fast magnetization dynamics in chemical processes, including biologically relevant pathways, as well as in materials science and technology.

This work was supported by the Hellenic Foundation for Research and Innovation (HFRI) and the General Secretariat for Research and Technology (GSRT), grant agreement No HFRI-FM17-3709 (project NUPOL) and by the project “HELLAS-CH” (MIS 5002735), which is implemented under the “Action for Strengthening Research and Innovation Infrastructures”, funded by the Operational Programme “Competitiveness, Entrepreneurship and Innovation” (NSRF 2014-2020) and cofinanced by Greece and the European Union (European Regional Development Fund).

\clearpage
\bibliography{ns_resolved_magnetometer_v2}
\bibliographystyle{unsrtnat}

\end{document}